\documentclass[conference]{IEEEtran}
\usepackage{graphicx}
\usepackage{multirow}

\usepackage{lineno}
\usepackage{amsfonts}
\ifCLASSINFOpdf
\else
\fi

\usepackage{caption}
\usepackage{algorithm}
\usepackage{algorithmic}
\usepackage{amstext}
\usepackage{graphicx}
\usepackage{amsmath}
\usepackage{amsfonts}
\usepackage{color}
\usepackage{array}
\usepackage[table,xcdraw]{xcolor}
\usepackage{float}
\usepackage[numbers,sort&compress,square]{natbib}

\usepackage{nomencl}
\makenomenclature
\usepackage{subcaption}
\begin{document}

\title{Smart Inverter Impacts on California Distribution Feeders with Increasing PV Penetration: A Case Study}

% author names and affiliations
% use a multiple column layout for up to three different
% affiliations
\author{

\IEEEauthorblockN{Zachary K. Pecenak}
\IEEEauthorblockA{Center for Energy Research \\
University of California San Diego\\
La Jolla, California 92093\\
Email: zpecenak@ucsd.edu}
\and
\IEEEauthorblockN{Jan Kleissl}
\IEEEauthorblockA{Jacobs school of Engineering \\
University of California San Diego\\
La Jolla, California 92093\\
Email: jkleissl@ucsd.edu}
\and
\IEEEauthorblockN{Vahid Rasouli Disfani}
\IEEEauthorblockA{Center for Energy Research \\
University of California San Diego\\
La Jolla, California 92093\\
Email: vrdisfani@ucsd.edu}

}
% conference papers do not typically use \thanks and this command
% is locked out in conference mode. If really needed, such as for
% the acknowledgment of grants, issue a \IEEEoverridecommandlockouts
% after \documentclass

% for over three affiliations, or if they all won't fit within the width
% of the page, use this alternative format:
%
%\author{\IEEEauthorblockN{Michael Shell\IEEEauthorrefmark{1},
%Homer Simpson\IEEEauthorrefmark{2},
%James Kirk\IEEEauthorrefmark{3},
%Montgomery Scott\IEEEauthorrefmark{3} and
%Eldon Tyrell\IEEEauthorrefmark{4}}
%\IEEEauthorblockA{\IEEEauthorrefmark{1}School of Electrical and Computer Engineering\\
%Georgia Institute of Technology,
%Atlanta, Georgia 30332--0250\\ Email: see http://www.michaelshell.org/contact.html}
%\IEEEauthorblockA{\IEEEauthorrefmark{2}Twentieth Century Fox, Springfield, USA\\
%Email: homer@thesimpsons.com}
%\IEEEauthorblockA{\IEEEauthorrefmark{3}Starfleet Academy, San Francisco, California 96678-2391\\
%Telephone: (800) 555--1212, Fax: (888) 555--1212}
%\IEEEauthorblockA{\IEEEauthorrefmark{4}Tyrell Inc., 123 Replicant Street, Los Angeles, California 90210--4321}}

% use for special paper notices
%\IEEEspecialpapernotice{(Invited Paper)}

% make the title area
\maketitle

% As a general rule, do not put math, special symbols or citations
% in the abstract
\begin{abstract}
The impacts of high PV penetration on distribution feeders have been well documented within the last decade. To mitigate these impacts, interconnection standards have been amended to allow PV inverters to regulate voltage locally. However, there is a deficiency of literature discussing how these inverters will behave on real feeders under increasing PV penetration. In this paper, we simulate several deployment scenarios of these inverters on a real California distribution feeder. We show that minimum and maximum voltage, tap operations, and voltage variability are improved due to the inverters. Line losses were shown to increase at high PV penetrations as a side effect. Furthermore, we find inverter sizing was shown to be important as PV penetration increased. Finally we show that increasing the number of inverters and removing the deadband from the Volt/VAr control curve improves the effectiveness.
\end{abstract}

% no keywords

% For peer review papers, you can put extra information on the cover
% page as needed:
% \ifCLASSOPTIONpeerreview
% \begin{center} \bfseries EDICS Category: 3-BBND \end{center}
% \fi
%
% For peerreview papers, this IEEEtran command inserts a page break and
% creates the second title. It will be ignored for other modes.
\IEEEpeerreviewmaketitle

\section{Introduction}
\label{intro}
The negative impacts of high PV penetration on electrical grids are well documented \cite{elnozahy2013technical}. In distribution systems, rooftop PV can cause large voltage fluctuations, over-voltages, and increased activation and maintenance costs for voltage regulators. Over-voltages exceeding the mandated ANSI C84.1 voltage limits of $\pm 5\%$ \cite{standardc84} pose a threat to behind-the-meter equipment. These problems become increasingly severe with increasing PV penetration \cite{nguyen2016high}.

In order to combat the negative effects of PV, both IEEE 1547 \cite{1547} and CA rule 21 \cite{rule21} interconnection standards have been amended to allow the local regulation of voltage through real and reactive power modulation. Given that PV inverters already function to regulate the real power injected into the grid from the PV, they are a natural choice to function as the control agent. These inverters which act to modulate electrical bus real and reactive power (as well as several other advanced functions \cite{seal2013common}), are hereafter referred to smart inverters (SI).

Power factor control (PFC) is one control method in which the power factor of the inverter operated on a fixed or variable schedule to regulate voltage. Volt/VAr control (VVC) is one of the functionalities of SI which follows some predetermined Volt/VAr curves. Typically VVC employs a negative sloped curve that instructs reactive power injection for low voltage and reactive power absorption for high voltage. Typical Volt/VAr curves are shown in figure \ref{VVCcurve}.

\cite{smith2011smart} introduces 7 types of inverter voltage control schemes, 3 of which are PFC while the other 4 are VVC. Through a simple simulation they conclude that VVC improves voltage conditions for PV penetrations of 20\%. \cite{reno2013smart} showed that increased voltages and voltage fluctuations due to PV on a feeder can be damped by both PFC and VVC, with VVC being more effective.

\cite{seuss2015improving} investigated the impact of local VVC on hosting capacity, which they define as the upper limit of PV penetration that does not cause violations in the network operating standards (i.e. ANSI range A limits). Their simulations on 6 feeders show that VVC can improve feeder hosting capacity, where the effectiveness increases with increasing inverter size.

\cite{abate2015smart} simulated a feeder with a number of VVC settings to determine which settings are the most appropriate. The authors quantify improvements in voltage, tap operations, line losses, and voltage fluctuations in terms of droop control settings and substation voltage.  However, the one PV penetration considered was not specified in the paper. The authors conclude that optimal settings are feeder specific.

Most recently, \cite{PSCVV} simulated the J1 feeder using realistic load and PV generation data at 15-min resolution to look at effects of Volt/VAr control on feeder voltages. The authors show that SI using VVC are capable of reducing maximum voltages on the feeder due to increasing PV production. The PV penetration was not specified by the authors.

To date, literature is lacking any discussion of SI operating under extremely high PV penetrations (PV penetration higher than 50\%) where adverse effects are greatest. Furthermore, all works assume that 100\% of PV systems have VVC capabilities, which is unlikely considering already high PV penetrations in several states such as California and the high cost of retrofitting existing systems. In this work, we systematically quantify the SI effects at PV penetration up to 200\% and the effects of different fractions of SI. Voltage maxima and minima, line losses, tap operations, and voltage variability are quantified on a real California distribution feeder. Also, in light of the result of \cite{abate2015smart}, we use a VVC curve with and without a deadband to determine the effect on this feeder.

In section \ref{simulation}, we introduce the distribution feeder model, relevant definitions, and simulation parameters. In section \ref{VVC} we introduce the control scheme used for the SI. In section \ref{Res}, we discuss the results of quasi-steady state time series simulations. Finally we conclude the work in \ref{conclusions}.

\section{Methodology}
\subsection{simulation setup}
\label{simulation}

PV penetration is well accepted term in literature to describe the ratio of PV to load on a feeder. Here, PV penetration is defined as:
\begin{align}
PV_{pen}=PV_{{kVA}_{\rm Installed}}/load_{\rm peak}.
\label{Pv_pen}
\end{align}

$PV_{pen}$ is fixed for a given feeder configuration. To quantify the importance of solar power generation at a given time instant, we also define instantaneous PV penetration as
\begin{align}
PV_{Pen_{Instantaneous}}=PV_{kVA}(t_i)/load(t_i).
\end{align}

Here, we introduce the SI fraction ($Frac_{\rm SI}$), to describe the ratio of PV capacity that has SI functionality to the total amount of PV on the feeder
\begin{align}
Frac_{SI}=PV_{{kVA}_{\rm SI}}/PV_{{kVA}_{\rm total}}.
\end{align}

\begin{figure}[!h]
\centering
\includegraphics[width=\columnwidth,trim={65 220 65 220},clip]{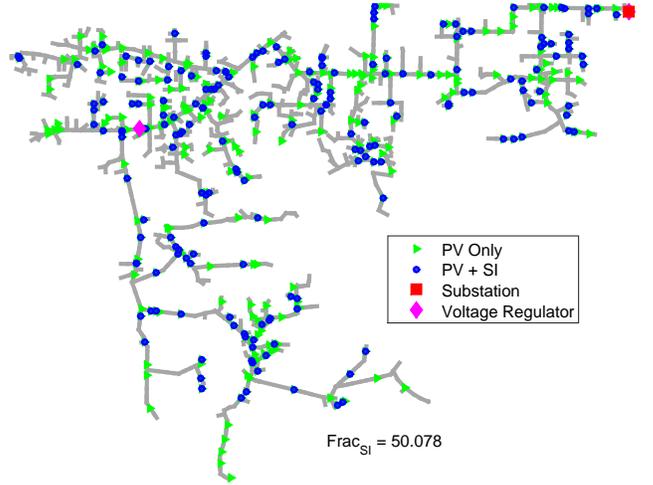}
\caption{California feeder with 50\% Smart Inverter fraction.}
\label{feeder}
\end{figure}

Figure \ref{feeder} visually depicts a SI fraction of 50\% on the distribution feeder simulated. The PV systems with SI were randomly selected in sequence until 49.5\% capacity was reached. If a selected PV system caused the total capacity to exceed 51\%, the entire process was repeated.\\
Since the feeder is long at 11.7~km to the furthest bus, voltage regulators exist at the substation and in the middle of the feeder. As a result, the substation voltage setpoint is set to 0.99 p.u.. The feeder is simulated with a total of 107 PV systems, with two of the systems being larger than 0.5~MW.

Individual generation profiles were created for each PV system. The profiles were generated using a sky imager following the methodology introduced in \cite{nguyen2016high}. A single time-series shape scaled according to the maximum power of the load was used for every load in the circuit.

Table \ref{QSTSTable} summarizes the simulation setup used in this work.
\begin{center}
\small
\vspace*{-\baselineskip}
\begin{table}[h]
\caption{QSTS simulation setup}
\begin{tabular}{ l|l }
Distribution Feeder & Rural California feeder\\
%\hline
Simulation Period&95 days (12/10/2014-3/15/2015)\\
PV Penetrations (\%)&0, 5, 10, 15, 25, 50, 75, 100, 125, 150, 175, 200 \\
SI Fraction (\%)&0, 50, 100\\
Evaluation Metrics&Voltage, line losses, tap operations, fluctuations\\
Simulation software&OpenDSS, Matlab\\
Load data&30 second, scaled substation load profile\\
PV data&30 second, sky image projections\\
\end{tabular}
\label{QSTSTable}
\end{table}
\end{center}

\subsection{Inverter Control }
\label{VVC}

\cite{seal2013common} introduces a handful of schemes for voltage regulation using SI. However the literature review in section \ref{intro} showed Volt/VAr control to be an extremely effective approach. Furthermore, VVC is desirable as it prioritizes active power output, meaning no active power is sacrificed to regulate voltage. As a consequence, the amount of inductive or capacitive VArs the inverter can use are restricted by the amount of active power being generated ($Q=\sqrt{S^2-P^2}$).

Because of this, the ratio of inverter AC power rating to the DC rating of the PV panel becomes important. Generally in practice, PV inverters are undersized (AC/DC ratio $\approx$ 0.8) to increase energy conversion efficiency. However, the largest PV impacts often coincide with high active power output. Therefore, undersized inverters will not be able to provide sufficient reactive power for voltage regulation. Although SI sizing has not been investigated in the literature, \cite{seal2013common} showed that an AC/DC ratio = 1.1 was sufficient to achieve a constant power factor of 0.9. In this work, all SI are slightly over-sized with an AC/DC ratio of 1.05.
\begin{figure}[!h]
\centering
\includegraphics[width=\columnwidth]{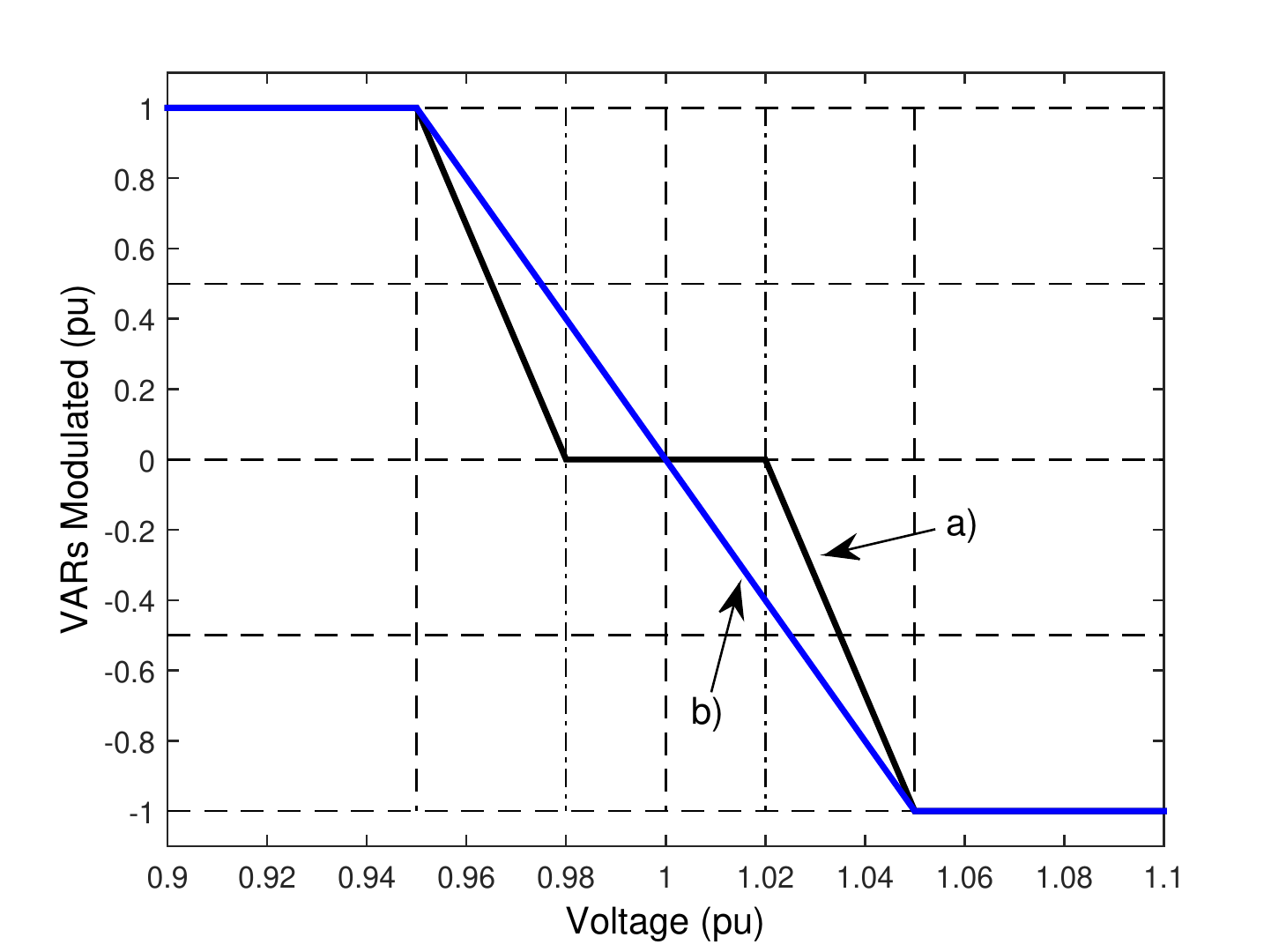}
\caption{Control curves used in QSTS. Curve a) has a deadband, while curve b) does not.}
\label{VVCcurve}
\end{figure}

Two VVC curves are investigated in this work, which are given in figure \ref{VVCcurve}. Curve a) has an low voltage limit of 0.95, where it produces the maximum amount of available capacitive VArs, and an upper limit of 1.05 where it produces the maximum amount of available inductive VArs. The curve also exhibits a deaband of 0.04 p.u. between 0.98 and 1.02 p.u., i.e. there is no reactive power support for voltages in that range. Curve b) has the same upper and lower voltage limits as curve a), but has no deadband.

\section{Results}
\label{Res}

\subsection{Voltage-Distance Profiles}
\label{VF}
\begin{figure}
\centering
	\begin{subfigure}[b]{0.50\textwidth}
	\includegraphics[width=\columnwidth]{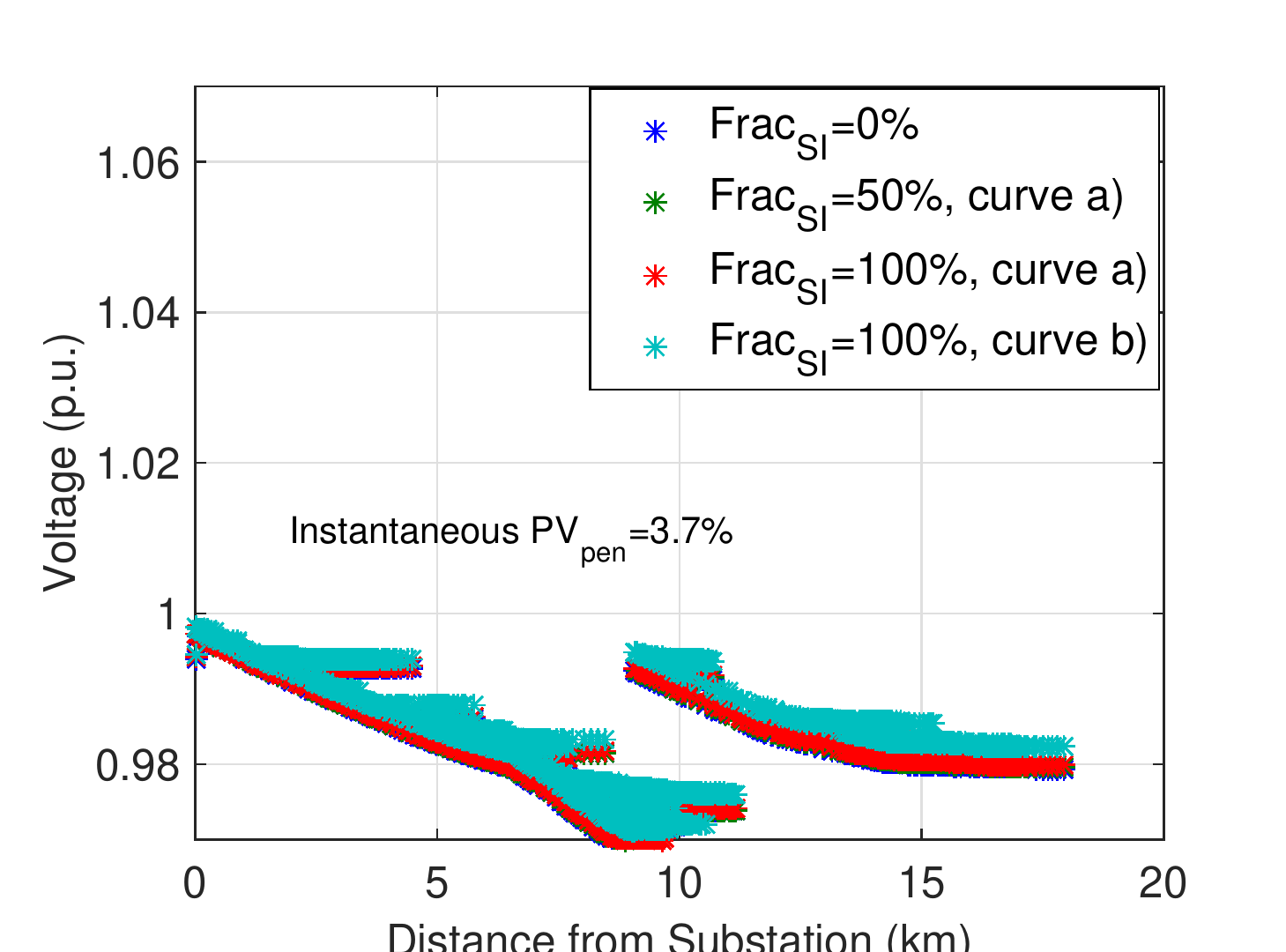}
	\caption{}
	\label{5Pen}
	\end{subfigure}

	\begin{subfigure}[b]{0.50\textwidth}
	\includegraphics[width=\columnwidth]{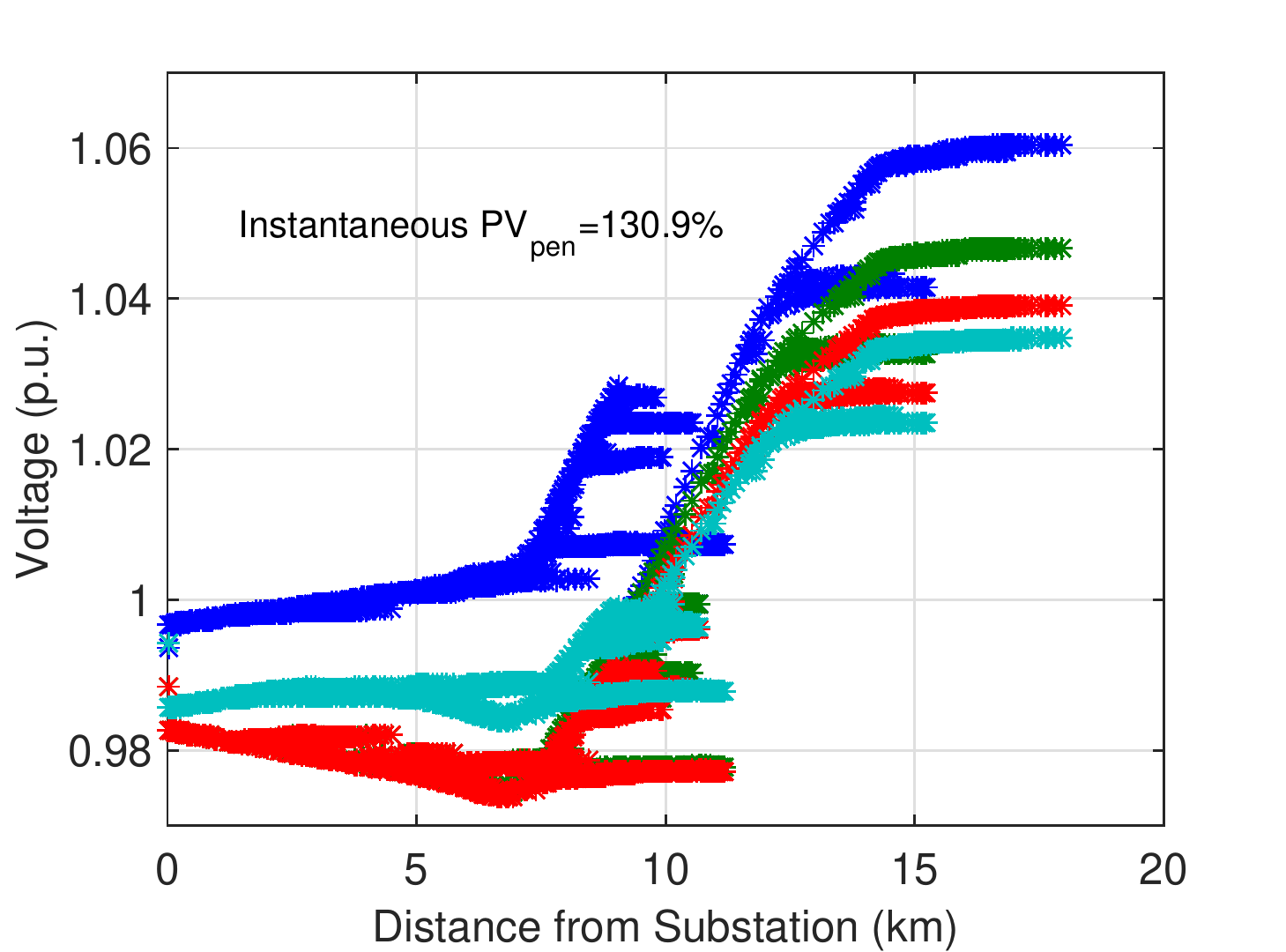}
	\caption{}
	\label{200Pen}
	\end{subfigure}

    \caption{Feeder voltage profile as a function of distance for 1/3/2015 12:00:00 PST. (a) 5\% nominal PV penetration. (b) 200\% nominal PV penetration. The curves in each plot represent different SI fractions and VVC curves simulated.}
\end{figure}

Without PV, or at low PV penetrations, voltage decays with distance away from the substation (Figure \ref{5Pen}). Even at such low PV penetration, SI are able to raise bus voltages and counteract voltage decay.

However, as PV penetration increases (Figure \ref{200Pen}), larger voltages are observed at the end of the feeder as a result of large active power outputs. This scenario is observed in figure \ref{200Pen} by the $Frac_{SI}=0$ case, where end of the feeder voltage exceeds 1.06 p.u.. For this day, all SI scenarios drop the end of the feeder voltage below the ANSI limit of 1.05 p.u. The effectiveness of the SI is seen to increase with increasing $Frac_{SI}$ and removal of the deadband. In fact, the curve without the deadband is capable of decreasing end of feeder voltage, while simultaneously increasing voltages near the substation, creating a flatter voltage profile than any of the other scenarios.

\subsection{Maximum Voltage}
\label{MaxVolt}
\begin{figure}[!h]
\centering
\includegraphics[width=\columnwidth]{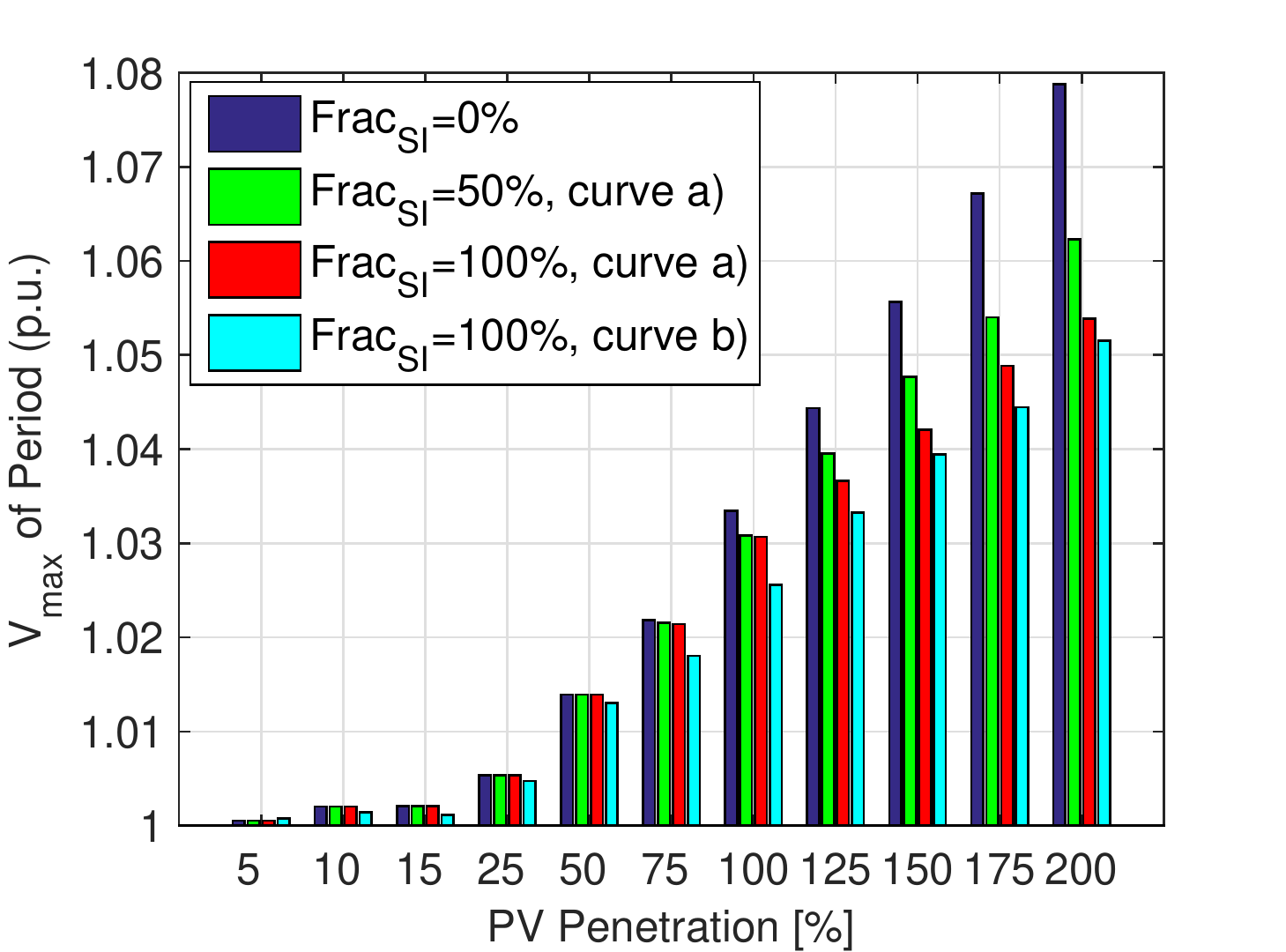}
\caption{Maximum voltage recorded on feeder during the 95 day simulation period. The maximum voltage is given for each SI scenario and each PV penetration.}
\label{VoltMax}
\end{figure}

The maximum voltage of the feeder over the 95 days simulation period was recorded and is given for each PV penetration in figure \ref{VoltMax}. In all cases the maximum voltage increases with increasing PV penetration, as expected. At all PV penetrations, the SI with VVC curve b) reduce the maximum voltage, whereas the SI operating under curve a) do not begin to have an effect until the minimum voltage is outside of the deadband range, which occurs first at $PV_{pen}=75\%$.

For PV penetrations greater than 150\%, the feeder without SI exceeds the ANSI 1.05 p.u. limit. All SI scenarios are able to mitigate the over-voltage condition at $PV_{\rm pen}=150\%$. However, $Frac_{\rm SI}=50\%$ exhibits violations at $PV_{\rm pen}>150\%$, while both $Frac_{\rm SI}=100\%$ scenarios experience violations at $PV_{\rm pen}-200\%$. The reason that the SI are not able to mitigate the over-voltage is due to the fact that VVC gives active power priority as discussed in section \ref{VVC}. Since the maximum voltage occurs when PV generates near its DC rating, the SI does not have sufficient VAr capacity to effectively mitigate the overvoltages.

\subsection{Minimum Voltage}
\label{MinVolt}
\begin{figure}[!h]
\centering
\includegraphics[width=\columnwidth]{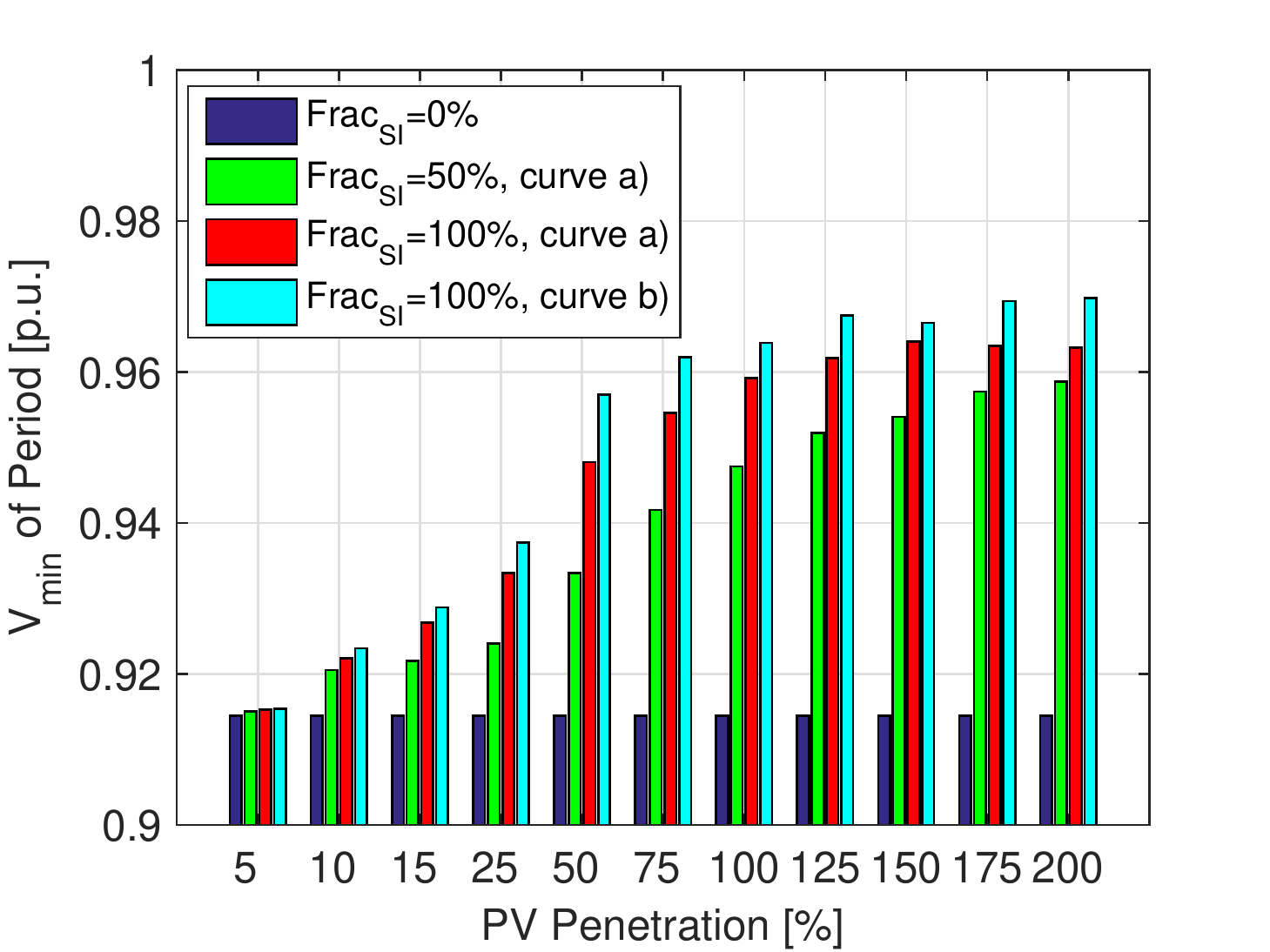}
\caption{Minimum voltage recorded on feeder during 95 day simulation period. The minimum voltage is given for each SI scenario and each corresponding PV penetration.}
\label{VoltMin}
\end{figure}

The voltage of the feeder recorded over the 95 days experiences a minimum of 0.916 p.u.. Since the minimum voltage occurs during a period of high load and zero PV output, it is not a function of PV penetration. However, as observed in Fig. \ref{VoltMin}, the effectiveness of SI to mitigate the under-voltage increases with increasing PV penetration. As PV penetration increases, the aggregate amount of available VArs on the feeder increases, thus giving more control of the voltage. Higher $Frac_{\rm SI}$ and the removal of the deadband lead to the highest minimum feeder voltages.

\subsection{Tap Operations}
\begin{figure}[!h]
\centering
\includegraphics[width=\columnwidth]{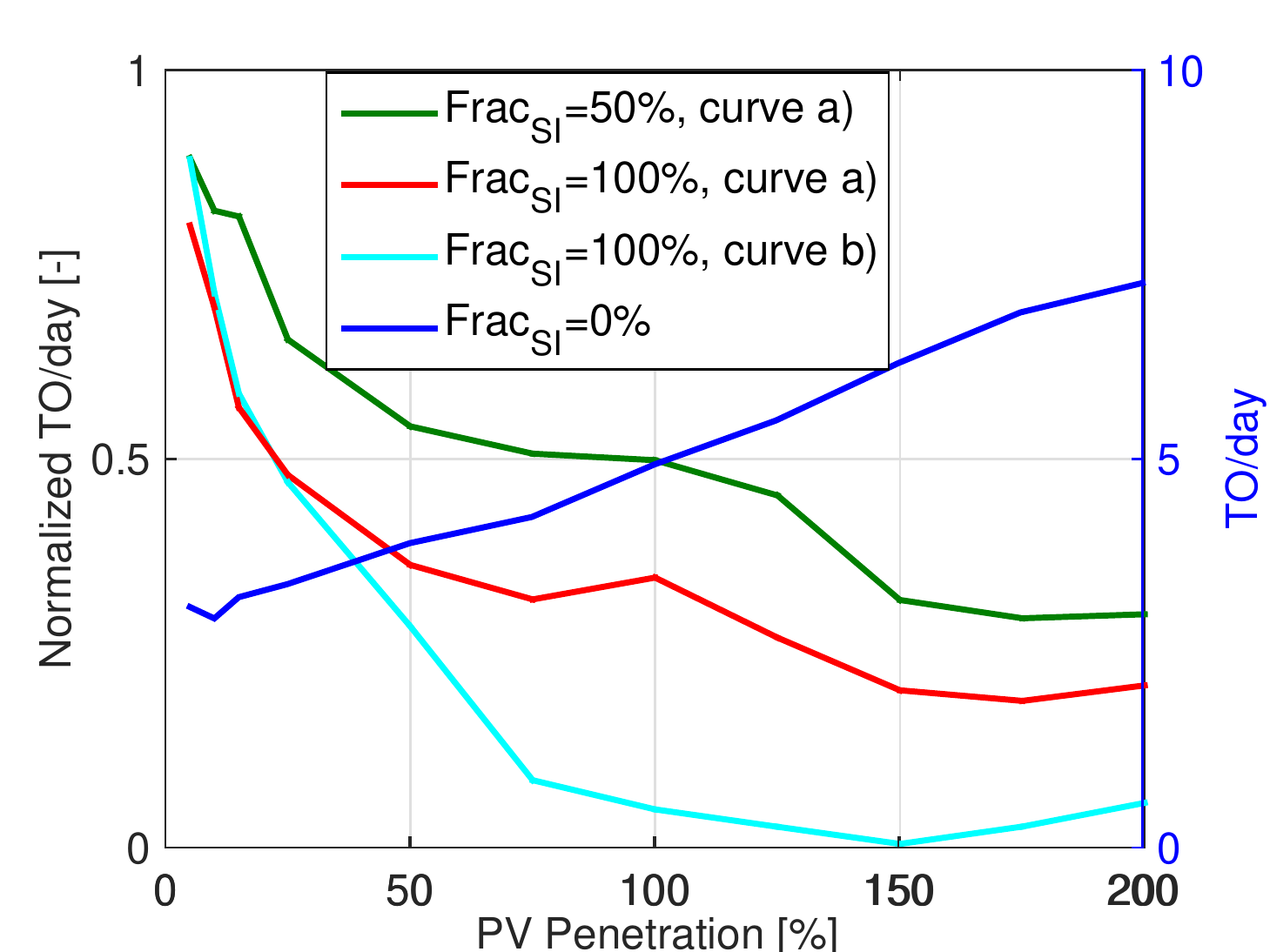}
\caption{Tap operations as a function of PV penetration. The plot has a left and right $y$-axis. The right sided $y$-axis gives the average number of tap operations for the feeder without SI as a function of PV penetration. The left sided axis shows the normalized tap operations on the feeder for the different SI cases, where the normalization is performed with respect to the case without SI.}
\label{TapOp}
\end{figure}

The right $y$-axis of Fig. \ref{TapOp} shows that on this feeder the number of tap operations increases from 3.1 tap operations at 5\% PV penetration to 7.3 tap operations per day at 200\% PV penetration. The left $y$ axis indicates that the greatest reduction in tap operations occurs for increasing $Frac_{SI}$ and removal of the deadband, but all SI scenarios reduce the tap operations substantially. In fact, for $frac_{SI}=100\%$ and VVC curve b), the SI reduce the tap operations per day to zero. It is expected this trend would continue for higher $PV_{pen}$, if the SI were over-sized more to allow for sufficient VAr support. This result supports the notion that SI -- if properly sited -- could act as primary voltage regulators on the circuit.
\label{TO}

\subsection{Line Losses}
\label{ll}
\begin{figure}[!h]
\centering
\includegraphics[width=\columnwidth]{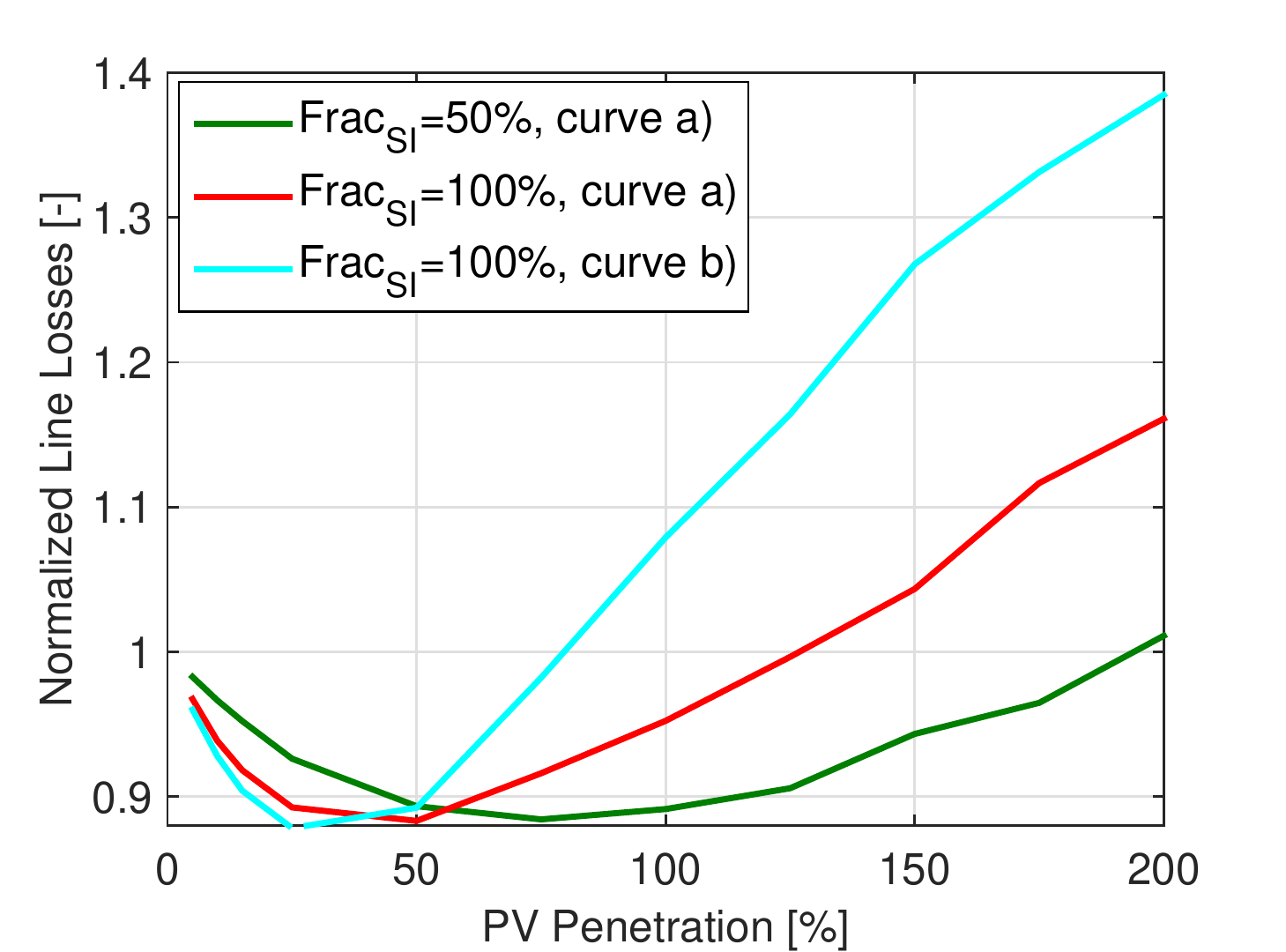}
\caption{Normalized power losses in feeder lines as a function of PV penetration. for each SI scenario. The normalization is with respect to the feeder without SI.}
\label{LineLoss}
\end{figure}

The normalized line losses in the feeder are given in figure \ref{LineLoss}. Initially ($PV_{pen}<25\%$) line losses are reduced for all SI scenarios. However, with increasing PV penetration all SI scenarios eventually produce greater line losses than the feeder without SI. The knee of the curve occurs at $PV_{pen}=25\%, 50\%$, and $75\%$ and the feeder starts producing line losses greater than the no SI case at $PV_{pen}=75\%, PV_{pen}=125\%$, and $PV_{pen}=190\%$ for $Frac_{SI}=100\%$ curve b), $Frac_{SI}=100\%$ curve a), and $Frac_{SI}=50\%$, respectively. These results indicate that the increase in line losses is proportional to the effectiveness of the SI control scheme.

At a high level the reason for increased line losses can be explained by the fact that the SI mitigate over-voltage through inductive VAr support. Since the voltage at the bus then decreases, the current in the line increases and the $I^2R$ losses increase.

\subsection{Voltage Fluctuations}
\label{VF}
Voltage fluctuations are often distinguished in two separate types, which have varying effects. Small fluctuations induce power-line flicker manifested through dimming and brightening of lights, but may be too small, localized, or short-lived to impact tap operations. Larger fluctuations cause tap/capacitor operation and can harm behind-the-meter equipment. To capture both types of voltage fluctuations, the solar variability characterization method of \cite{lave2015characterizing} is extended to measure voltage fluctuations.

\begin{figure}[!h]
\centering
\includegraphics[width=\columnwidth]{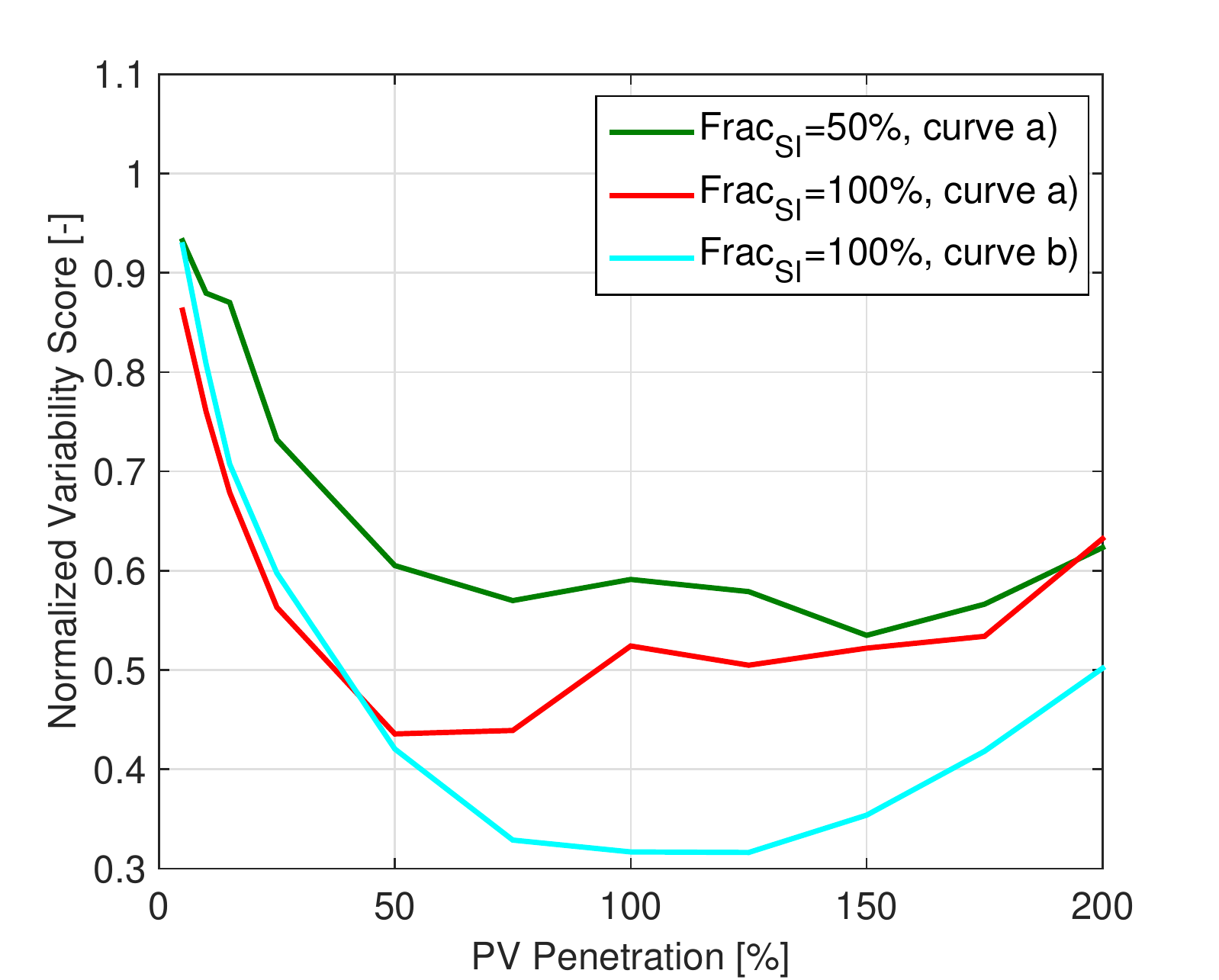}
\caption{Normalized VS of each SI scenario as a function of PV penetration.}
\label{VS}
\end{figure}

The idea is to assign a variability score (VS) to each feeder configuration over the 95 days. Essentially, the method weights the probability of fluctuations by magnitude, with respect to a set threshold, $V_0$. The greater the fluctuations relative to the limit, the higher the VS assigned to the feeder configuration.
\begin{align}
\widetilde{V}_{\Delta t}=1/{\Delta t} ( \sum_t^{t+\Delta t}V-\sum_{t-\Delta t}^t V)
\end{align}

\begin{align}
VS_{\widetilde{V}}(\Delta t)=100 \rm max[V_0 \times P(|\widetilde{V}_{\Delta t}|>V_0])
\end{align}
The VS is calculated by generating a cumulative distribution function (CDF) of voltage fluctuations $P(|\widetilde{V}_{\Delta t}|>V_0$ across all buses for the 95 simulation days. % This is repeated for each PV penetration, SI fraction, and feeder simulated.
Figure \ref{VS} plots the VS for each SI configuration normalized by the VS of the case with 0\% SI. For all SI scenarios and PV penetrations, the variability decreases compared to the case without SI. As observed in the other metrics, the effectiveness of the SI increases with increasing $Frac_{SI}$ and by removing the deadband. Furthermore, it is observed that the SI improve the VS more and more up to $PV_{pen}=125\%$, at which point the effectiveness decreases.

\section{Conclusions}
\label{conclusions}
In this work we have simulated a distribution feeder with increasing PV penetration. Varying amounts of smart inverters (SI) operating under different Volt/VAr control schemes were simulated to quantify the effect of SI on the feeder under increasing PV penetration. By observing the effect of SI on voltage (as a function of distance, maximum, and minimum), tap operations, line losses, and fluctuations the following conclusions were drawn.

\begin{itemize}
\item SI effectiveness increases with increasing $Frac_{SI}$.
\item Linear Volt/VAr control outperforms the deadband curve.
\item Line losses increase with increasing improvements in voltage.
\item AC/DC ratio is the limiting factor for voltage support.
\end{itemize}

% conference papers do not normally have an appendix

% use section* for acknowledgment
\section*{Acknowledgment}

The authors would like to thank the team at SunSpec and Matt Rylander of EPRI for the expertise they lent on this work. This work was funded by CEC grant PON-14-303.

% trigger a \newpage just before the given reference
% number - used to balance the columns on the last page
% adjust value as needed - may need to be readjusted if
% the document is modified later
%\IEEEtriggeratref{8}
% The "triggered" command can be changed if desired:
%\IEEEtriggercmd{\enlargethispage{-5in}}

% references section

% can use a bibliography generated by BibTeX as a .bbl file
% BibTeX documentation can be easily obtained at:
% http://mirror.ctan.org/biblio/bibtex/contrib/doc/
% The IEEEtran BibTeX style support page is at:
% http://www.michaelshell.org/tex/ieeetran/bibtex/
%\bibliographystyle{IEEEtran}
% argument is your BibTeX string definitions and bibliography database(s)
%\bibliography{IEEEabrv,../bib/paper}
%
% <OR> manually copy in the resultant .bbl file
% set second argument of \begin to the number of references
\bibliographystyle{IEEEtran}
\bibliography{bib}

% that's all folks
\end{document}